\documentclass[onecolumn,aps,12pt]{revtex4}
\usepackage{amsmath,graphicx}
\usepackage{feynmp}
\def\,{\ifmmode\mskip\thinmuskip\else\leavevmode\thinspace\fi}

\newcommand{\Ree}{\mbox{Re}}

\newcommand\ba{\begin{eqnarray}}
\newcommand\ea{\end{eqnarray}}
\def\Li#1#2{{\mathrm{Li}}_{#1}\left(#2\right)}
\newcommand{\nn}{\nonumber}

\begin{document}
\title{Process $e^+ e^- \to 3\pi (\gamma)$ with final state radiative corrections}

\author{S.~Bakmaev}
\email{bakmaev@theor.jinr.ru}
\affiliation{Joint Institute for Nuclear Research, 141980 Dubna,
Russia}

\author{Yu.~M.~Bystritskiy}
\email{bystr@theor.jinr.ru}
\affiliation{Joint Institute for Nuclear Research, 141980 Dubna,
Russia}

\author{E.~A.~Kuraev}
\email{kuraev@theor.jinr.ru}
\affiliation{Joint Institute for Nuclear Research, 141980 Dubna,
Russia}

\date{\today}% It is always \today, today,

\begin{abstract}
We consider the process of annihilation of
$e^+ e^-$ to three pion final state for the case of
moderately high energies. The final state emission of
virtual and real photon is considered explicitly.
The calculations are performed in frames
of QED with point-like mesons and pion part of
chiral perturbation theory.
Some numerical estimates are given.
\end{abstract}

%\pacs{Valid PACS appear here}
\maketitle

\section{Introduction}
The problem of taking into account of radiative corrections (RC) of
lowest order of perturbation theory to process of three pion production
in annihilation channel of colliding $e^+ e^-$ at moderately high energies
becomes urgent for precision measurement of hadronic contribution to
muon anomalous magnetic moment $(g-2)_\mu$ \cite{Bennett:2004pv}.
This is the motivation of this paper. The similar calculation for
production of 2 $\pi$ and $\mu^+ \mu^-$ at the annihilation channel was
performed recently \cite{Bystritskiy:2005ib}.

We shall use the pion sector of chiral perturbation theory (ChPT) to perform
the calculations of interaction of pions with electromagnetic field.
This theory is given by Wess-Zumino-Witten effective lagrangian
\cite{Witten:1983tx,Wess:1971yu}. The relevant piece of this lagrangian
is reproduced below:
\ba
    {\cal L} &=&
    \frac{f_\pi^2}{4}
    Sp\left[
        D_\mu U \left(D^\mu U\right)^+ +
        \chi U^+ + U \chi^+
    \right]-
    \nn \\
    &-&
    \frac{e}{16\pi^2} \varepsilon^{\mu\nu\alpha\beta}
    A_\mu
    Sp \left[
        Q\left\{
            \left(\partial_\nu U\right)
            \left(\partial_\alpha U^+\right)
            \left(\partial_\beta U\right)
            U^+
            -
            \left(\partial_\nu U^+\right)
            \left(\partial_\alpha U\right)
            \left(\partial_\beta U^+\right)
            U
        \right\}
    \right]-
    \nn \\
    &-&
    \frac{i e^2}{8\pi^2} \varepsilon^{\mu\nu\alpha\beta}
    \left(\partial_\mu A_\nu\right)
    A_\alpha
    Sp \left[
        Q^2 \left(\partial_\beta U\right) U^+ +
        Q^2 U^+ \left(\partial_\beta U\right) +
    \right.
    \nn \\
    &&\qquad\qquad\qquad\qquad\qquad +
    \left.
    \frac{1}{2} Q U Q U^+ \left(\partial_\beta U\right) U^+ -
    \frac{1}{2} Q U^+ Q U \left(\partial_\beta U^+\right) U
    \right],
    \label{ChiralLagrangian}
\ea
where $f_\pi = 94$ MeV is the pion decay constant,
$U = exp\left(i\frac{\sqrt{2} \Phi}{f_\pi}\right)$,
$D_\mu U = \partial_\mu U + i e A_\mu \left[Q, U\right]$,
$Q = diag\left(\frac{2}{3},-\frac{1}{3},-\frac{1}{3}\right)$
is the quark charge matrix and terms with
$\chi = Bdiag\left(m_u, m_d, m_s\right)$ introduce explicit chiral
symmetry breaking due to nonzero quark masses. The constant B has
dimension of mass and is determined through the equation
$B m_q = M^2$, where $m_q = m_u \approx m_d$ and $M$ is the pion mass.
The pseudoscalar meson matrix $\Phi$ has its standard form:
\ba
    \Phi =
    \left(
        \begin{array}{ccc}
            \frac{1}{\sqrt{2}}\pi^0 + \frac{1}{\sqrt{6}} \eta &
            \pi^+ &
            K^+ \\
            \pi^- &
            -\frac{1}{\sqrt{2}}\pi^0 + \frac{1}{\sqrt{6}} \eta &
            K^0 \\
            K^- &
            \bar K^0 &
            -\frac{2}{\sqrt{6}} \eta
        \end{array}
    \right).
\ea

% ------------------------------------------------------------------------------------
\section{Born approximation}
% ------------------------------------------------------------------------------------

\begin{figure}
\begin{fmffile}{BornFD}
\begin{fmfgraph*}(150,80)
\fmfstraight
\fmfleft{em,ep}
\fmf{fermion,tension=2}{em,fee,ep}
\fmf{photon,tension=2,label=$q$}{fee,fppp}
\fmf{dashes}{pip,fppp,pim}
\fmf{dashes}{fppp,pin}
\fmfright{pip,pim}
\fmfright{pin}
\fmflabel{$e^-(p_-)$}{em}
\fmflabel{$e^+(p_+)$}{ep}
\fmflabel{$\pi^-(q_-)$}{pim}
\fmflabel{$\pi^+(q_+)$}{pip}
\fmflabel{$\pi^0(q_0)$}{pin}
\end{fmfgraph*}
\end{fmffile}
\vspace{0cm}
\caption{Feynman diagram contributing the process probability in Born approximation.}
\label{BornFD}
\end{figure}
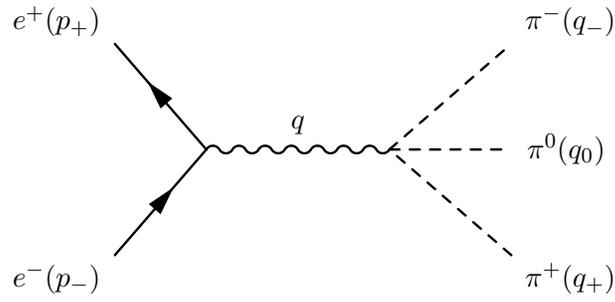

We consider reaction of $e^+~e^-$ annihilation into three pions:
\ba
    e^-(p_-) + e^+(p_+) \to
    \pi^-(q_-) + \pi^+(q_+) + \pi^0(q_0).
\ea
Matrix element for this process in Born approximation have a form (see Fig. \ref{BornFD}):
\ba
M^{(0)} = \frac{i \alpha}{\pi f_{\pi}^{(0) 3}}\cdot\frac{1}{q^2} \cdot \bar{v}(p_+)\gamma_{\mu} u(p_-) \cdot (\mu q_+ q_- q_0),
\ea
where $f_{\pi}^{(0)}$ is the unrenormalized pion decay constant, $s = q^2 = (p_+ + p_-)^2$ -- invariant mass of initial state,
$(\mu q_+ q_- q_0) \equiv \varepsilon_{\mu\nu\alpha\beta} q_+^\nu q_-^\alpha q_0^\beta$.

Squaring this matrix element and performing summation over initial leptons spin states we
get the total cross section for this process:
\ba
    \sigma_B^{(0)} = \frac{\alpha^2 s^2}{2^8\cdot3 \cdot \pi^5 f_{\pi}^{(0) 6}}
    \int^{x_+^{max}}_{x_+^{min}} dx_+
    \int^{x_-^{max}}_{x_-^{min}} dx_-~
    G(x_+,x_-),
    \label{Born}
\ea
here
\ba
    G(x_+,x_-) = 4 (x_+^2-\mu^2)(x_-^2 -\mu^2) - \left(1 - 2 x_+ - 2 x_- +2 x_+ x_- + \mu^2\right)^2,
\ea
with $\mu^2 = M^2/s$, $M$ -- is pion mass, $x_\pm = \varepsilon_\pm/\sqrt{s}$,
$x_0 = \varepsilon_0/\sqrt{s}$ -- are fractions of final pion's energies,
while $x_+ + x_- + x_0 = 1$. Limits of integration in (\ref{Born}) follows from kinematical constrains:
\ba
    x_+^{min} &=& \mu, \qquad
    x_+^{max} = \frac{1}{2}\left(1 - \frac{3M^2}{s}\right) = \frac{1}{2}\left(1 - 3 \mu^2\right), \nn\\
    x_-^{max, min} &=& \frac{1}{2(1-2 x_+ + \mu^2)}\left((1-x_+)(1-2 x_+ + \mu^2) \pm R(x_+)\right),    \nn
\ea
where $R^2(x_+) = (x_+^2 - \mu^2)(1-2 x_+ + \mu^2)(1-2 x_+ - 3\mu^2)$.

% ------------------------------------------------------------------------------------
\section{Virtual photon emission}
% ------------------------------------------------------------------------------------

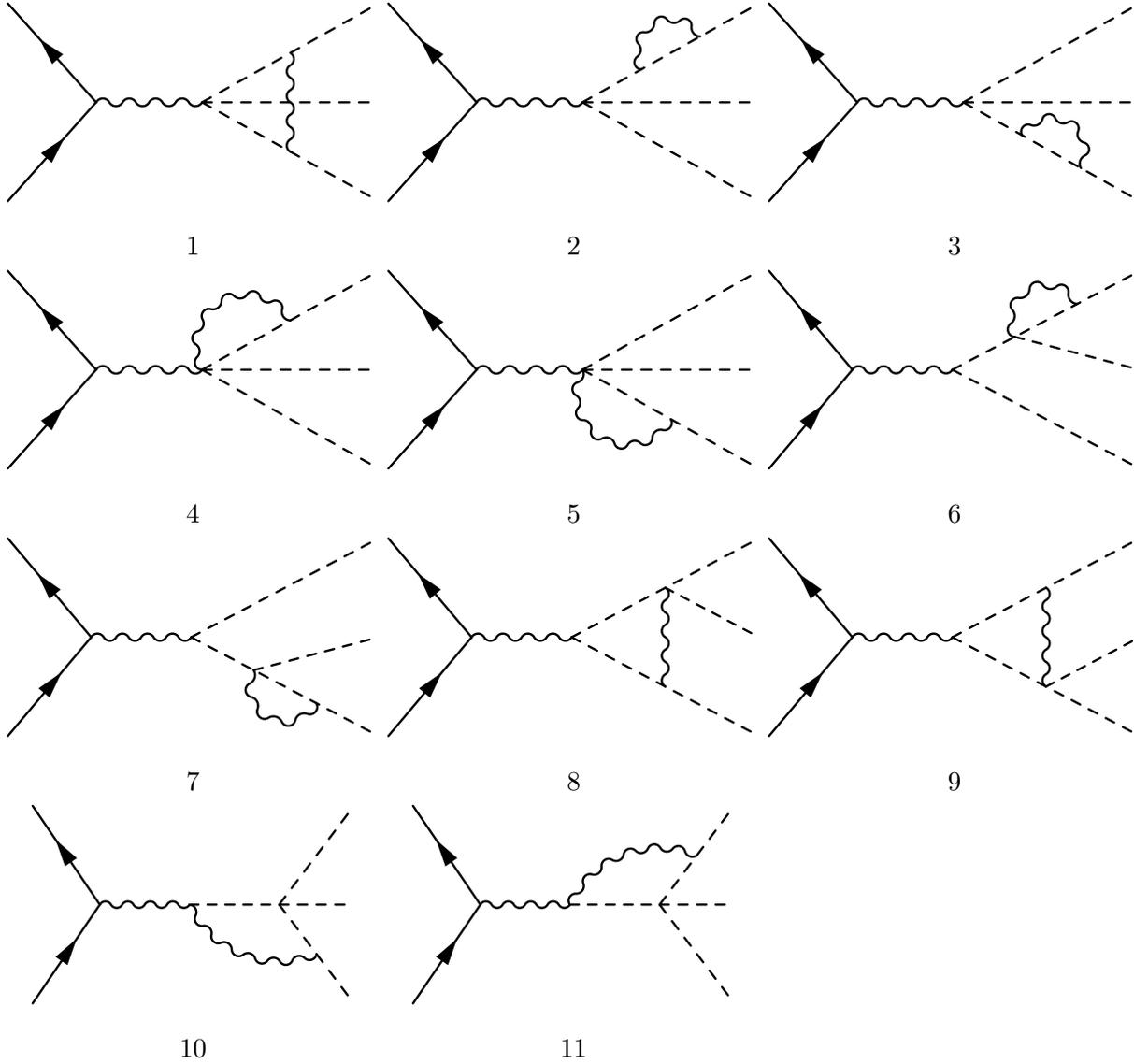
\begin{figure}
\begin{tabular}{ccc}
% ---------------------------------------------
%  Diagram 1.
% ---------------------------------------------
\begin{fmfgraph*}(150,80)
\fmfstraight
\fmfleft{em,ep}
\fmf{fermion,tension=1.8}{em,fee,ep}
\fmf{photon,tension=3}{fee,fppp}
\fmf{dashes,tension=0.6}{fppp,pip}
\fmf{dashes,tension=0.6}{fppp,pim}
\fmf{dashes,tension=0.6}{fppp,pin}
%\fmfright{pip,pim,pin}
\fmfright{pip,pin,pim}
\fmffreeze
\fmf{phantom}{pip,v1,fppp,v2,pim}
\fmffreeze
\fmf{photon}{v1,v2}
\end{fmfgraph*}
&
% ---------------------------------------------
%  Diagram 2.
% ---------------------------------------------
\begin{fmfgraph*}(150,80)
\fmfstraight
\fmfleft{em,ep}
\fmf{fermion,tension=1.8}{em,fee,ep}
\fmf{photon,tension=3}{fee,fppp}
\fmf{dashes,tension=0.6}{fppp,pip}
\fmf{dashes,tension=0.6}{fppp,pim}
\fmf{dashes,tension=0.6}{fppp,pin}
\fmfright{pip,pin,pim}
\fmffreeze
\fmf{phantom}{fppp,v1,v2,pim}
\fmffreeze
\fmf{photon,left,tension=0}{v1,v2}
\end{fmfgraph*}
&
% ---------------------------------------------
%  Diagram 3.
% ---------------------------------------------
\begin{fmfgraph*}(150,80)
\fmfstraight
\fmfleft{em,ep}
\fmf{fermion,tension=1.8}{em,fee,ep}
\fmf{photon,tension=3}{fee,fppp}
\fmf{dashes,tension=0.6}{fppp,pip}
\fmf{dashes,tension=0.6}{fppp,pim}
\fmf{dashes,tension=0.6}{fppp,pin}
%\fmfright{pip,pim,pin}
\fmfright{pip,pin,pim}
\fmffreeze
\fmf{phantom}{fppp,v1,v2,pip}
\fmffreeze
\fmf{photon,left,tension=0}{v1,v2}
\end{fmfgraph*}
\\
1 & 2 & 3\\
% ---------------------------------------------
%  Diagram 4.
% ---------------------------------------------
\begin{fmfgraph*}(150,80)
\fmfstraight
\fmfleft{em,ep}
\fmf{fermion,tension=1.8}{em,fee,ep}
\fmf{photon,tension=3}{fee,fppp}
\fmf{dashes,tension=0.6}{fppp,pip}
\fmf{dashes,tension=0.6}{fppp,pim}
\fmf{dashes,tension=0.6}{fppp,pin}
%\fmfright{pip,pim,pin}
\fmfright{pip,pin,pim}
\fmffreeze
\fmf{phantom}{fppp,v1,pim}
\fmffreeze
\fmf{photon,left,tension=0}{fppp,v1}
\end{fmfgraph*}
&
% ---------------------------------------------
%  Diagram 5.
% ---------------------------------------------
\begin{fmfgraph*}(150,80)
\fmfstraight
\fmfleft{em,ep}
\fmf{fermion,tension=1.8}{em,fee,ep}
\fmf{photon,tension=3}{fee,fppp}
\fmf{dashes,tension=0.6}{fppp,pip}
\fmf{dashes,tension=0.6}{fppp,pim}
\fmf{dashes,tension=0.6}{fppp,pin}
\fmfright{pip,pin,pim}
\fmffreeze
\fmf{phantom}{fppp,v1,pip}
\fmffreeze
\fmf{photon,right,tension=0}{fppp,v1}
\end{fmfgraph*}
&
% ---------------------------------------------
%  Diagram 6.
% ---------------------------------------------
\begin{fmfgraph*}(150,80)
\fmfstraight
\fmfleft{em,ep}
\fmf{fermion,tension=1.8}{em,fee,ep}
\fmf{photon,tension=3}{fee,fppp}
\fmf{dashes,tension=0.8}{fppp,pip}
\fmf{dashes,tension=0.8}{fppp,pim}
\fmfright{pip,pin,pim}
\fmffreeze
\fmf{phantom}{fppp,v1,v2,pim}
\fmffreeze
\fmf{dashes}{v1,pin}
\fmf{photon,left,tension=0}{v1,v2}
\end{fmfgraph*}
\\
4 & 5 & 6\\
% ---------------------------------------------
%  Diagram 7.
% ---------------------------------------------
\begin{fmfgraph*}(150,80)
\fmfstraight
\fmfleft{em,ep}
\fmf{fermion,tension=1.8}{em,fee,ep}
\fmf{photon,tension=3}{fee,fppp}
\fmf{dashes,tension=0.8}{fppp,pip}
\fmf{dashes,tension=0.8}{fppp,pim}
\fmfright{pip,pin,pim}
\fmffreeze
\fmf{phantom}{fppp,v1,v2,pip}
\fmffreeze
\fmf{dashes}{v1,pin}
\fmf{photon,right,tension=0}{v1,v2}
\end{fmfgraph*}
&
% ---------------------------------------------
%  Diagram 8.
% ---------------------------------------------
\begin{fmfgraph*}(150,80)
\fmfstraight
\fmfleft{em,ep}
\fmf{fermion,tension=1.8}{em,fee,ep}
\fmf{photon,tension=3}{fee,fppp}
\fmf{dashes,tension=0.8}{fppp,pip}
\fmf{dashes,tension=0.8}{fppp,pim}
\fmfright{pip,pin,pim}
\fmffreeze
\fmf{phantom}{pip,v1,fppp,v2,pim}
\fmffreeze
\fmf{photon}{v1,v2}
\fmf{dashes}{v2,pin}
\end{fmfgraph*}
&
% ---------------------------------------------
%  Diagram 9.
% ---------------------------------------------
\begin{fmfgraph*}(150,80)
\fmfstraight
\fmfleft{em,ep}
\fmf{fermion,tension=1.8}{em,fee,ep}
\fmf{photon,tension=3}{fee,fppp}
\fmf{dashes,tension=0.8}{fppp,pip}
\fmf{dashes,tension=0.8}{fppp,pim}
\fmfright{pip,pin,pim}
\fmffreeze
\fmf{phantom}{pip,v1,fppp,v2,pim}
\fmffreeze
\fmf{photon}{v1,v2}
\fmf{dashes}{v1,pin}
\end{fmfgraph*}
\\
7 & 8 & 9\\
% ---------------------------------------------
%  Diagram 10.
% ---------------------------------------------
\begin{fmfgraph*}(130,80)
\fmfstraight
\fmfleft{em,ep}
\fmf{fermion,tension=1}{em,fee,ep}
\fmf{photon,tension=1.5}{fee,v}
\fmf{dashes,tension=1.5}{v,fppp}
\fmf{dashes,tension=0.6}{fppp,pip}
\fmf{dashes,tension=0.6}{fppp,pim}
\fmf{dashes,tension=0.6}{fppp,pin}
\fmfright{pip,pin,pim}
\fmffreeze
\fmf{phantom}{pip,v1,fppp,v2,pim}
\fmffreeze
\fmf{photon,right=0.4,tension=0}{v,v1}
\end{fmfgraph*}
&
% ---------------------------------------------
%  Diagram 11.
% ---------------------------------------------
\begin{fmfgraph*}(130,80)
\fmfstraight
\fmfleft{em,ep}
\fmf{fermion,tension=1}{em,fee,ep}
\fmf{photon,tension=1.5}{fee,v}
\fmf{dashes,tension=1.5}{v,fppp}
\fmf{dashes,tension=0.6}{fppp,pip}
\fmf{dashes,tension=0.6}{fppp,pim}
\fmf{dashes,tension=0.6}{fppp,pin}
\fmfright{pip,pin,pim}
\fmffreeze
\fmf{phantom}{pip,v1,fppp,v2,pim}
\fmffreeze
\fmf{photon,left=0.4,tension=0}{v,v2}
\end{fmfgraph*}
& \\
10 & 11 &\\
\end{tabular}
\caption{Feynman diagrams of emission of additional virtual photon.}
\label{VirtualRCsFD}
%\end{table}
\end{figure}

Radiative corrections from emission of virtual photon can be represented
by 11 Feynman diagrams (FD) (see Fig. \ref{VirtualRCsFD}).

First we shall notice that FDs 4, 5, 6, 7, 10, 11 gives zero contribution
($\delta_4 = \delta_5 = \delta_6 = \delta_7 = \delta_{10} = \delta_{11} = 0$)
due to
\ba
    \int\frac{d^4k}{i\pi^2}
    \frac{(\mu k q (2 q_- - k))}
    {k^2 (k^2 - 2kq_-) (k^2 - 2kq + q^2 - M^2)}
    \equiv 0.
\ea
FDs 2 and 3 are the contributions from pion wave function renormalization and they
are equal to \cite{Chang}:
\ba
    \delta_c = \delta_2 + \delta_3 = \frac{\alpha}{\pi}
    \left(L_\Lambda + \ln\frac{M^2}{\lambda^2} - \frac{3}{4}\right),
\ea
where $L_\Lambda = \ln(\Lambda^2/M^2)$ and $\Lambda$ -- is the ultraviolet cut-off parameter,
$\lambda$ -- is the fictitious photon mass.
Considering contributions of FDs 1, 8, 9 we get:
\ba
    \delta_v &=& \delta_1 + \delta_8 + \delta_9 = \frac{\alpha}{\pi}
    \left[
        1 + \frac{1}{2} L_\Lambda
        -
        \frac{1+\beta^2}{2\beta} \ln\left(\frac{1+\beta}{1-\beta}\right)
        \ln\frac{M^2}{\lambda^2}
        +
    \right.
    \nn \\
    && \qquad\qquad\qquad\qquad
    \left.
        +
        \frac{1+\beta^2}{4} s_1~
        \Ree \int^1_0 \frac{dx}{q_x^2}
        \left(
            \ln\frac{q_x^2}{M^2} - 2
        \right)
        + Q
    \right],
    \label{VirtualRC189}
    \\
    Q &=&
        \int^1_0 dx
        \int^1_0 dy y
        \ln\frac{-x(1-x y) + x x_+(1-y) + \mu^2 y}{-x(1-x y) + x x_-(1-y) + \mu^2 y}. \label{Q}
\ea
Here $s_1 = (q_+ + q_-)^2$ is the invariant mass of charged pions pair,
$\beta^2 = 1 - 4M^2/s_1$,
$q_x^2 = M^2 - s_1 x(1-x) - i 0$.
The integrals in (\ref{VirtualRC189}) can be calculated explicitly:
\ba
    \Ree \int^1_0 \frac{dx s_1}{q_x^2} &=& -\frac{2}{\beta} L, \\
    \Ree \int^1_0 \frac{dx s_2}{q_x^2} \ln\frac{q_x^2}{M^2} &=&
    \frac{4}{\beta}
    \left[
        L \ln\left(\frac{1+\beta}{2\beta}\right)
        -
        \frac{1}{4} L^2
        +
        \Li{2}{\frac{1-\beta}{1+\beta}}
        +
        2\xi_2
    \right],
\ea
where $L = \ln\frac{1+\beta}{1-\beta}$, $\Li{2}{x} = -\int_0^x\frac{dt}{t}\ln(1-t)$ and $\xi_2 = \frac{\pi^2}{6}$.

% ------------------------------------------------------------------------------------
\section{Soft real photon emission}
% ------------------------------------------------------------------------------------

The standard calculation of contribution of real soft photon emission
by final pions
\ba
    \delta_s = \frac{\sigma_{soft}}{\sigma_B} =
    -\frac{\alpha}{4\pi^2}
    \int \frac{d^3k}{\omega}
    \left.
    \left(
        \frac{q_-}{k q_-} - \frac{q_+}{k q_+}
    \right)^2
    \right|_{\omega < \Delta \varepsilon},
\ea
where $\Delta \varepsilon$ is the maximum energy of soft photon (i.e. $\omega < \Delta\varepsilon$),
leads to:
\ba
    \delta_s &=& \frac{2\alpha}{\pi}
    \left\{
        \left(
            \ln\Delta - \frac{1}{2}\ln(x_+ x_-) + \ln\frac{M}{\lambda}
        \right)
        \left(
            -1
            +
            \frac{1+\beta^2}{2\beta}
            L
        \right)
        +
    \right. \nn \\
    && \qquad
        +
        \left.
        \frac{1+\beta^2}{4\beta}
        \left[
            -g - \frac{1}{2}L^2
            +L
             \ln\left(\frac{4}{1-\beta^2}\right)
            -\xi_2
            -2\Li{2}{-\frac{1-\beta}{1+\beta}}
        \right]
    \right\},
    \label{Soft}
\ea
where $\Delta = \Delta\varepsilon/\sqrt{s}$ and the quantity $g$ is defined by
\ba
    g = 2\beta \int_0^1 \frac{dt}{1-\beta^2t^2}
    \ln
    \left(
        1 +
        \frac{1-t^2}{4}\frac{\left(x_+ - x_-\right)^2}{x_+ x_-}
    \right).
    \label{g}
\ea

% ------------------------------------------------------------------------------------
\section{Hard real photon emission}
% ------------------------------------------------------------------------------------

Consider the contribution of radiative corrections
which arises from the emission of additional hard photon by final particles, i.e. the process:
\ba
    e^-(p_-) + e^+(p_+) \to
    \pi^-(q_-) + \pi^+(q_+) + \pi^0(q_0) + \gamma(k).
\ea
Amplitude for this process can be written in the form:
\ba
    M = (4\pi \alpha)^2\frac{1}{q^2} \cdot \bar{v}(p_+)\gamma_{\mu} u(p_-) \cdot
    \frac{1}{4\pi^2 f_\pi^3} \cdot T_{\mu\nu} e^\nu(k),
\ea
where $k$, $e_\mu(k)$ are the momenta and the polarization vector of final real photon.
$T_{\mu\nu}$ is the tensor corresponding to the
$\gamma^*(\mu,Q) \to \pi^+(q_+) \pi^-(q_-) \pi^0(q_0) \gamma(\nu,k)$ vertex,
which follows from (\ref{ChiralLagrangian}):
\ba
    T^{\mu\nu} &=& (\mu\nu Q k)A + (\mu\nu(Q+k)q_0) +
        (\mu \lambda Q q_0)
        \left(
            \frac{q_-^\nu q_+^\lambda}{(q_- k)}
            +
            \frac{q_+^\nu q_-^\lambda}{(q_+ k)}
        \right)
        - \nn \\
    &&
        -
        (\nu \lambda k q_0)
        \left(
            \frac{(2 q_- - Q)^\mu q_+^\lambda}{Q^2 - 2 (q_- Q)}
            +
            \frac{(2 q_+ - Q)^\mu q_-^\lambda}{Q^2 - 2 (q_+ Q)}
        \right),
        \label{TensorT}
\ea
where $Q = q_+ + q_- + q_0 + k$, (thus $Q_0 = \sqrt{s}$),
$A = 1 - (s_1-M^2)/((Q-k)^2-M^2)$.
This tensor satisfies the gauge-invariance condition for both photon legs:
\ba
    Q_\mu T^{\mu\nu} =
    k_\nu T^{\mu\nu} = 0.
\ea
Thus we may write the cross section in a form:
\ba
    \sigma^{hard} =
    \frac{(4\pi\alpha)^3}{8s} \cdot
    \frac{1}{s^2}
    Sp\left[\hat p_+ \gamma^\mu \hat p_- \gamma^\nu \right]
    \left(\frac{1}{4\pi^2 f_\pi^3}\right)^2
    \frac{1}{3}
    \left(
        g_{\mu \nu} - \frac{Q_\mu Q_\nu}{Q^2}
    \right)
    \int d\Gamma_4 \sum_\lambda
    \left|
        T^{\alpha\beta} e^\lambda_\beta
    \right|^2.
    \label{HardCrossSection1}
\ea
Phase volume for final state has a form:
\ba
    d\Gamma_4 &=& (2\pi)^4
    \delta^4
    \left(
        p_+ + p_- - q_+ - q_- - q_0 - k
    \right)
    \frac{d^3q_+}{(2\pi)^3 2\varepsilon_+}
    \frac{d^3q_-}{(2\pi)^3 2\varepsilon_-}
    \frac{d^3q_0}{(2\pi)^3 2\varepsilon_0}
    \frac{d^3k}{(2\pi)^3 2 \omega} =
    \nn \\
    &=&
    (2\pi)^{-8} \frac{s^2 \pi^2}{16}
    x~dx~dx_+~dx_-~dO_\gamma,
\ea
where $x = \omega/\sqrt{s}$,~$x+x_++x_-+x_0 = 1$.
And now the cross section takes the form:
\ba
    \sigma^{hard} =
    \frac{\alpha^3}{2^8 \cdot 3 \cdot \pi^7 f_\pi^6}
    \int dx~x~dx_+~dx_-~dO_\gamma
    \left.
    \left(
        -\sum_\lambda \left| T^{\mu\nu} \right|^2
    \right)\right|_{x > \Delta}.
\ea
We shall notice, that the sum of hard photon emission
$\delta_h = \sigma^{hard}/\sigma_B$
and soft real photon emission $\delta_s$ contributions ($\delta_h+\delta_s$)
does not contain the auxiliary
parameter $\Delta$. To see this explicitly let us consider the
small $x=\omega/\sqrt{s}$ limit of $\sigma^{hard}$. Really if we consider the case
$\Delta \sqrt{s} < \omega \ll \sqrt{s}$ then (see (\ref{TensorT})):
\ba
    \left.
    T^{\mu\nu} e_\nu(k)
    \right|_{\omega \ll \sqrt{s}}
    \approx
        (\mu q_+ q_- q_0)
        \left(
            \frac{(q_- e)}{(q_- k)}
            -
            \frac{(q_+ e)}{(q_+ k)}
        \right).
\ea
We can calculate the hard photon emission contribution in this limit:
\ba
    \int \frac{d^3k}{2\pi \omega} \sum_\lambda
    \left.\left(
        -\left| T^{\mu\nu} e_\nu(k) \right|^2
    \right)\right|_{\omega \to \Delta\sqrt{s}}
    =
    4 \ln \frac{1}{\Delta}
    \left(
        \frac{1+\beta^2}{2\beta} \ln\left(\frac{1+\beta}{1-\beta}\right)
        -1
    \right),
\ea
and we get for $\delta_h$:
\ba
    \left.
    \delta_h
    \right|_{\omega \to \Delta\sqrt{s}}
    \approx
    \frac{2\alpha}{\pi}
    \left[
    \ln \frac{1}{\Delta}
    \left(
        \frac{1+\beta^2}{2\beta} \ln\left(\frac{1+\beta}{1-\beta}\right)
        -1
    \right) + O(\Delta)
    \right].
\ea
We redefine the contributions in the following manner:
\ba
    \delta_s + \delta_h \to \bar\delta_s + \bar\delta_h,
\ea
\ba
    \bar\delta_s &=& \delta_s +
    \frac{2\alpha}{\pi}
    \ln \frac{1}{\Delta}
    \left(
        \frac{1+\beta^2}{2\beta} \ln\left(\frac{1+\beta}{1-\beta}\right)
        -1
    \right),
    \\
    \bar\delta_h &=& \delta_h -
    \frac{2\alpha}{\pi}
    \ln \frac{1}{\Delta}
    \left(
        \frac{1+\beta^2}{2\beta} \ln\left(\frac{1+\beta}{1-\beta}\right)
        -1
    \right),
\ea
where both $\bar\delta_s$ and $\bar\delta_h$ are not dependent on $\Delta$ any more.

% ------------------------------------------------------------------------------------
\section{Conclusion}
% ------------------------------------------------------------------------------------
The final result is
\ba
    \sigma^{ee\to3\pi(\gamma)}
    &=&
    \sigma_B^{(0)}
    \left(1 + \delta_{sw}\right)
    \left(1 + \delta\right)
    =
    \sigma_B
    \left(1 + \delta\right),
    \qquad
    \sigma_B = \sigma_B^{(0)}\left(f_\pi^{(0)} \to f_\pi\right)
    \\
    \delta &=&
    \left.
    \left(
        \delta_c
        +
        \delta_v
        +
        \bar\delta_s
        +
        \bar\delta_h
    \right)
    \right|_{L_\Lambda = 0},
    \qquad
    \delta_{sw} = \frac{3\alpha}{2\pi} L_\Lambda,
\ea
where we extracted the short-distance contributions in form
$\left(1 + \delta_{sw}\right)$ and used this factor to
renormalize $f_\pi^{(0)}$ pion decay constant in form
$f_\pi^{(0)-6}\left(1 + \delta_{sw}\right) = f_\pi^{-6}$ \cite{Holstein:ua}.

The explicit form of $\delta-\bar\delta_h$ is:
\ba
    \delta-\bar\delta_h
    &=&
    \frac{\alpha}{\pi}
    \left\{
        -\frac{1}{2}\ln\left(x_+ x_-\right)
        \left( -1 + \frac{1+\beta^2}{2\beta} L\right)
        +\frac{1}{4} + Q
    \right.
    +
    \nn \\
    &&\qquad
    +
    \frac{1+\beta^2}{4\beta}
    \left[
        -g - \frac{1}{2} L^2 + L \ln\frac{4}{1-\beta^2}
        -\xi_2 - 2 \Li{2}{-\frac{1-\beta}{1+\beta}}
    \right]
    +
    \nn \\
    &&\qquad
    +
    \left.
    \frac{1+\beta^2}{\beta}
    \left[
        L + L \ln\frac{1+\beta}{2\beta}
        -\frac{1}{4}L^2
        +\Li{2}{-\frac{1-\beta}{1+\beta}}
        +2\xi_2
    \right]
    \right\},
    \label{DeltaDeltah}
\ea
with $L=\ln\left(\frac{1+\beta}{1-\beta}\right)$, $g$ defined in (\ref{g}) and $Q$ defined in (\ref{Q}).
We note that $g(x_+ = x_-) = Q(x_+ = x_-) = 0$.
The form of $\bar\delta_h$ depends on
the experimental conditions of final state particles registration and
not considered here.

\begin{figure}
\includegraphics[scale=1]{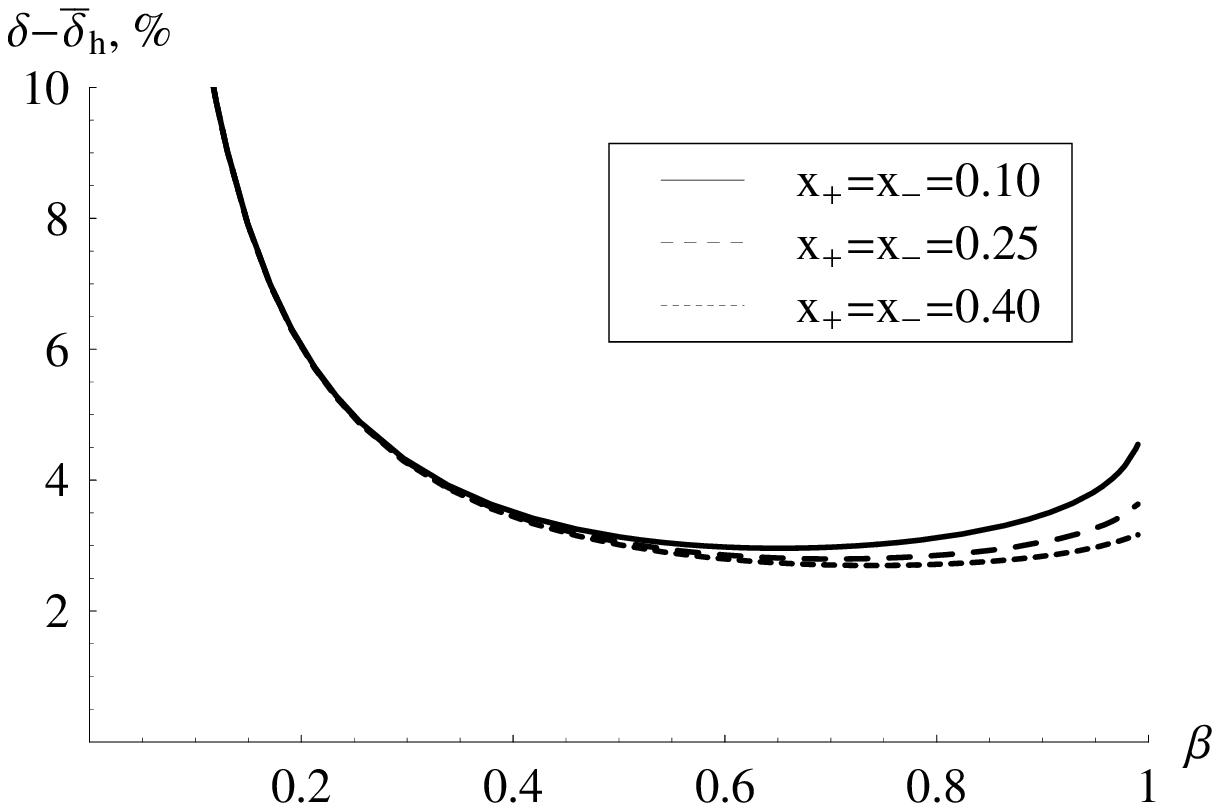}
\caption{The value $\delta-\bar\delta_h$ (see (\ref{DeltaDeltah})) in percents for
typical experimental conditions as function of $\beta$.}
\label{DeltaGraphics}
\end{figure}

In Fig. \ref{DeltaGraphics} we present the value $\delta-\bar\delta_h$ for typical experimental
situation.

% ------------------------------------------------------------------
\begin{acknowledgements}
% ------------------------------------------------------------------
We are grateful to G.~V.~Fedotovich for attracting our attention to
this problem. We are also grateful to
Z.~K.~Silagadze for valuable discussions.
\end{acknowledgements}

% ------------------------------------------------------------------------------------
%  The bibliography.
% ------------------------------------------------------------------------------------

% ------------------------------------------------------------------------------------
%  The diagram process in Born approximation.
% ------------------------------------------------------------------------------------

% ------------------------------------------------------------------------------------
%  The diagrams of virtual photon emission.
% ------------------------------------------------------------------------------------
\begin{fmffile}{VirtualRCs}
%\begin{table}[pp]
\end{fmffile}

% ------------------------------------------------------------------------------------
%  Graphic of total radiative corrections.
% ------------------------------------------------------------------------------------


\begin{thebibliography}{99}

%\cite{Bennett:2004pv}
\bibitem{Bennett:2004pv}
  G.~W.~Bennett {\it et al.}  [Muon g-2 Collaboration],
  %``Measurement of the negative muon anomalous magnetic moment to 0.7-ppm,''
  Phys.\ Rev.\ Lett.\  {\bf 92}, 161802 (2004)
  [arXiv:hep-ex/0401008].
  %%CITATION = HEP-EX 0401008;%%

%\cite{Bystritskiy:2005ib}
\bibitem{Bystritskiy:2005ib}
  Y.~M.~Bystritskiy, E.~A.~Kuraev, G.~V.~Fedotovich and F.~V.~Ignatov,
  %``The cross sections of the muons and charged pions pairs production at
  %electron positron annihilation near the threshold,''
  arXiv:hep-ph/0505236.
  %%CITATION = HEP-PH 0505236;%%

%\cite{Witten:1983tx}
\bibitem{Witten:1983tx}
  E.~Witten,
  %``Current Algebra, Baryons, And Quark Confinement,''
  Nucl.\ Phys.\ B {\bf 223}, 433 (1983).
  %%CITATION = NUPHA,B223,433;%%

%\cite{Wess:1971yu}
\bibitem{Wess:1971yu}
  J.~Wess and B.~Zumino,
  %``Consequences Of Anomalous Ward Identities,''
  Phys.\ Lett.\ B {\bf 37} (1971) 95.
  %%CITATION = PHLTA,B37,95;%%

%\cite{Chang}
\bibitem{Chang}
  Ngee-Pong~Chang,
  %``Electromagnetic Correction Effects on the $\pi^+ \to \pi^0 e^+ \nu$ Decay,''
  Phys.\ Rev.\ {\bf 131} (1963) 1272.
  %%CITATION = PHLTA,B37,95;%%

%\cite{Holstein:ua}
\bibitem{Holstein:ua}
B.~R.~Holstein,
% ``How Large Is F(Pi)?,''
Phys.\ Lett.\ B {\bf 244}, 83 (1990).
%%CITATION = PHLTA,B244,83;%%

\end{thebibliography}
\end{document}